\documentstyle[aps,prl,twocolumn,epsf]{revtex}

\begin{document}
\draft
\wideabs{
\title{Nonlinear Radiation Pressure and Stochasticity in
Ultraintense Laser Fields}
\author{Joel E. Moore}
\address{Department of Physics,
Massachusetts Institute of Technology, Cambridge, MA 02139}
\date{February 13, 1998}
\maketitle
\begin{abstract}
The drift acceleration due to radiation reaction for a single electron
in an ultraintense plane wave ($a = eE/mc\omega \sim 1$) of arbitrary
waveform and polarization is
calculated and shown to be proportional
to $a^3$ in the high-$a$ limit.  The cyclotron motion of
an electron in a constant
magnetic field and an ultraintense plane wave is numerically found
to be quasiperiodic even in the
high-$a$ limit if the magnetic field is not too
strong, as suggested by previous analytical work.
A strong magnetic field causes highly chaotic electron motion and
the boundary of the highly chaotic region of parameter space is determined
numerically and shown to agree with analytical predictions.
\end{abstract}

\pacs{PACS numbers: 52.40.Nk 52.35.Mw}

}
It has been known for many years that qualitative changes occur in the
behavior of an electron moving in an electromagnetic plane wave when the
dimensionless strength $a = e E / m c \omega$ is of order unity.
Recent advances in
laser pulse compression and amplification~\cite{perry}
have made it possible to attain
such ultraintense waves in the
laboratory and led to new investigations of their properties.
One important effect is that an electron moving in an
ultraintense electromagnetic plane wave and also subjected to
slowly varying ``background'' fields behaves approximately like a particle of
enhanced mass
$m \gamma_0 = m \sqrt{1 + e^2 \langle A_\mu A^\mu \rangle / m^2 c^4}$
drifting in the background fields.  Here $A$ is the vector potential
and $\gamma_0 = \sqrt{1 + a^2/2}$ for a linearly polarized monochromatic
wave.  The fast motion
in the wave can be rigorously averaged over if the background fields
are sufficiently weak and the plane wave is not so strong that
pair creation effects become significant.

The first part of this Letter uses the guiding-center equations
obtained in the derivation of the enhanced-mass approximation~\cite{moore}
to calculate the drift acceleration of an electron in a plane wave
due to the radiation damping and reaction forces.  In a strong
wave, the variation of the electron effective mass $m \gamma$
over the wave orbit must be properly accounted for to obtain the correct
drift acceleration.  The average drift acceleration is
the quantity of greatest interest because experiments and
astrophysical phenomena typically
involve many wave periods.  The guiding-center equations provide
the tools to carry out the averaging and determine the motion
of electrons under the (previously calculated) Lorentz-Dirac
radiation force~\cite{landau,sarachik}.
The second part of this Letter considers
the destruction of the enhanced-mass behavior and transition to
stochasticity as the strength
of the background fields is increased to violate the field-strength condition
$e B_{\rm back} / m c \omega_{\rm wave} \ll 1$ necessary for the
enhanced-mass derivation.  The breakdown of the enhanced-mass picture
is shown numerically to predict the onset of stochasticity even for very strong
wave intensity.  Both the nonlinear radiation acceleration and the
stochasticity are expected to be significant in astrophysical
situations, and the first effect may well be visible in laboratory
experiments with ultraintense laser pulses.

Classical calculations are valid for high
$a$ as long as the discrete photon energy $\hbar \omega \ll m c^2$ and
the amplitude for QED processes such as $e^+ e^-$ pair
creation is small ($e E \hbar / mc \ll mc^2$)~\cite{hartemann}.
The classical high-$a$ regime includes a number of existing and proposed
accelerator designs, such as
the plasma wakefield and beat-wave accelerators~\cite{tajima}.
Lasers used in
the National Ignition Facility and similar inertial fusion projects
have $a \sim 1$ so strong-field effects are significant.  Many
astrophysical problems also involve high-$a$ radiation
sources, and in particular consequences of strong-wave radiation
damping are discussed in~\cite{kegel}.

The motion of an electron in a plane wave of arbitrary intensity is
integrable both classically and within the Dirac equation.  This
occurs because a third conserved quantity $\gamma - p_x$
exists in addition to the two
transverse generalized momenta.  Here $\gamma$ is the electron Lorentz
factor and $p_x$ is the component of electron momentum along the wave axis.
Including radiative effects or adding
additional fields generally violates this conservation.
The result of the scattering of the wave by the electron
is a force on the electron along the wave axis, which has
the form $F = 2 e^4 \langle {\bf E}^2 \rangle /
3 m^2 c^4$ for weak plane waves.
In the high-$a$ regime the electron radiates much more strongly
($\propto a^4$ rather than $a^2$ in the low-$a$ limit) and radiates
high harmonics of the wave frequency~\cite{landau,sarachik}.

Previous work on the motion caused by radiation reaction in a
strong wave has included numerical studies~\cite{grewing,hartemann}
and solutions to the
Lorentz-Dirac equation (in the Landau approximation~\cite{landau})
in special cases such as monochromatic linearly and
circularly polarized waves~\cite{anders}. The
guiding-center formalism gives a simple result for the drift
velocity, the quantity of primary interest, valid for an arbitrarily
polarized, polychromatic wave.
The two approximations involved are the Landau approximation that
the radiation force is calculated from the trajectory in the absence
of the radiation force, and the guiding-center approximation
that the changes in the drift velocity over a single orbit
are nonrelativistic and hence add linearly, which is well satisfied
for radiation reaction in the classical regime.  An additional
advantage of the guiding-center formalism is that multiple applied forces
effectively superpose, which is not in general the case
for nonlinear equations of motion.

The motion in a wave along ${\bf \hat x}$
with dimensionless vector potential ${\bf A}(t - x/c)= {\bf A}(\eta)$
is given in the ``drift frame''
where the electron has zero average velocity
by $p_i = - m c A_i$ for the transverse components,
and $p_x = m c ({\bf A}^2 - \langle {\bf A}^2 \rangle) / 2 \gamma_0$, with
$\langle \rangle$ indicating averaging over $\eta$ and
${\gamma_0}^2 = 1 + \langle {\bf A}^2 \rangle.$
The wave envelope is assumed constant in $\eta$ so that the averages
are well-defined.  Variation in the wave envelope
causes a ponderomotive force, reviewed below.

The radiation reaction force from scattered photons causes the
drift velocity (the velocity of the drift frame) to change in time.
The drift
velocity component along the wave axis changes for an applied force ${\bf F}$
(in addition to the force from the wave)
according to~\cite{moore}
\begin{equation}
{d v^x_d \over dt} = {F_x \over m \gamma} -
{F_y v_y \over m c \gamma_0} - {F_z v_z \over m c \gamma_0}.
\label{gceq}
\end{equation}
Note that the
instantaneous drift acceleration from an applied force is not necessarily
in the direction of the applied force.
The causality problems with the Lorentz-Dirac
classical radiation force~\cite{hartemann} do not appear in the
Landau approximation, which is valid if the radiation force is a small
perturbation on the motion causing the radiation.
The instantaneous radiation force on the electron is
\begin{eqnarray}
F_i &=& {2 e^2 \over 3 m c^3} [{d \over dt} (\gamma {dp_i \over dt})
- {v_i \gamma^2 \over m c^2} (({d{\bf p} \over dt})^2
- ({dE \over c dt})^2)] \cr
&=& {2 e^2 {\gamma_0}^2 \over 3 m c^3} [{d^2 p_i \over d\eta^2}
- {v_i {\gamma_0}^2 \over m c^2} (({d{\bf p} \over d\eta})^2 -
({dE \over c d\eta})^2)].
\label{radforce}
\end{eqnarray}
Upon substituting (\ref{radforce}) into (\ref{gceq}), we obtain the
instantaneous acceleration (w.r.t. phase) in the x-direction:
${d v^x_d / d\eta} = {2 e^2 \gamma_0 ({d{\bf A} / d\eta})^2
/ 3 m c^2}.$
Now the averaging over $\eta$ is simple.  For a wave with intensity
per frequency interval $I(\omega)$, the result is
\begin{equation}
\langle {d v^x_d \over dt} \rangle =
{8 \pi e^4 \int I(\omega)\,d\omega \over 3 m^3 c^5}
\sqrt{1 + {4 \pi e^2 \over m^2 c^3} \int {I(\omega)\,d\omega \over \omega^2}}.
\end{equation}
The acceleration is thus independent of polarization and the phasing between
different frequences, just as in the nonrelativistic limit, but scales
with $a^3$ rather than $a^2$ for large $a$.

For a linearly polarized monochromatic wave, the average acceleration is
$2 e^2 \omega^2 (a^2 / 2 + a^2 / 4) \sqrt{1 + a^2 / 2} / 3 m c^2$. 
For circular polarization, the result is
$2 e^2 \omega^2 (a^2 + a^4) \sqrt{1 + a^2} / 3 m c^2$.  These are
essentially identical to the considerably less transparent
solutions in~\cite{anders}.  For circular
polarization this is just the radiated power divided by ${\bar m} c$,
with ${\bar m} = m \gamma_0$.  For linear polarization this is not the
case, but the correct acceleration is obtained if we account for the
fact that the electron's total radiated momentum is nonzero, unlike for
circular polarization~\cite{sarachik}.
The correct average radiation acceleration is
thus obtained from conservation of energy and momentum {\it if}
a drifting electron is assigned momentum $m \gamma_0 v_d$.  However,
there is no {\it a priori} reason for this assignment without the
use of (\ref{gceq}).

The average radiation force can be defined
as the required constant force to balance the above acceleration:
$F_{\rm rad} = m \gamma_0 A_{\rm rad}$.  The radiation force from a
blackbody source of intensity $I$ and temperature $T$ is
\begin{equation}
F_{\rm rad} = {8 \pi e^4 I \over 3 m^2 c^5} (1 + {10 \hbar^2 e^2 I \over
\pi m^2 c^3 k_B^2 T^2}).
\end{equation}

Changes in the envelope of a plane wave cause a
ponderomotive force, which conserves $\gamma - p_x$ is
directed forward when the wave is rising and backward when the wave is
falling.
(There are also transverse ponderomotive forces in real beams arising
from the finite spot size; these have a different character and are
not discussed here.)
The radiation force does not conserve $\gamma - p_x$ and always acts
in the forward direction.  Treating an electron in a plane wave as an
enhanced-mass
particle acted upon by ponderomotive and radiation forces gives
a simple and accurate description of single-particle behavior
in the classical regime.

Now we turn to consider the second problem mentioned in the opening:
the destruction of the enhanced-mass picture
when strong constant electromagnetic fields
are added to the plane wave.  It will be shown that for one typical
field configuration, i.e., one without special symmetries giving rise
to integrability, the breakdown of the enhanced-mass picture is consistent
with the analytical predictions of~\cite{moore} and associated with
the onset of stochasticity over a wide range of beam intensity.
An example of an integrable configuration is an applied magnetic field
parallel to the wave axis~\cite{roberts}.
In the following the wave will
be taken to have constant amplitude and the radiation force will be
neglected.
We study a constant magnetic field
${\bf B} = B {\bf \hat z}$ perpendicular to a linearly polarized
wave $A_y(\xi)$
traveling in the ${\bf \hat x}$ direction.
Then $p_z$ is constant and we take $p_z = 0$ so that the electron motion
is confined to the $xy$ plane.
For small wave strength $a_w = e E / m c \omega$ the motion
can be analyzed perturbatively because the equations of motion
are nearly linear~\cite{karney}.

\begin{figure}
\epsfxsize=2.0truein
\centerline{\epsffile{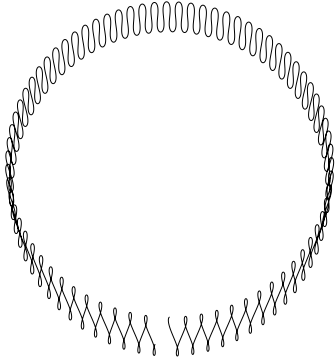}}
\caption{Typical electron cyclotron motion with $a_w = 1.5$, $a_b = 0.02$,
${\bf B} \parallel {\bf \hat z}$ and ${\bf k} \parallel {\bf \hat x}$.}
\label{figone}
\end{figure}

When $a_w$ is of order unity, the equations are strongly nonlinear
and new phenomena appear.   However, the motion is still simple for
$a_w > 1$ as long as the applied magnetic field is not too strong.
In the derivation of the equations for
the motion of the guiding-center, it was necessary to assume
$a_b = e B / m c \omega = \omega_c / \omega \ll 1$, i.e., the electron
is far from resonance.  Fig.~\ref{figone} shows a typical trajectory
in the enhanced-mass regime: the electron executes
fast oscillations in the wave, while its guiding-center makes a slow
orbit in the magnetic field.  The relativistic nonlinearity in the
equations of motion can destroy quasiperiodicity, but only
for $\omega_c / \omega \sim 1$. 

With no wave present
the gyrocenter of the electron motion
$(x_0,y_0) = (x + {p_y / e B}, y - {p_x / e B})$
(with $m = c = 1$) is constant in time.
The
electron still has an exactly constant gyrocenter for an arbitrarily
strong wave:
\begin{equation}
(K_x, K_y) = (x + {p_y + e A_y(\xi) \over e B}, y - {p_x + 1
- \gamma \over
e B}).
\label{relgyro}
\end{equation}
Such a gyrocenter exists
for any direction of the magnetic field $B_0 {\bf \hat b}$
and any polarization of the
wave:
${\bf r}_c = {\bf r} - ({\bf r} \cdot {\bf \hat b}) {\bf \hat b} -
	{{\bf \hat b} \times ({\bf p} + e {\bf A}(\xi) + 1 - \gamma {\bf
\hat k}) / e B_0}$.

Now consider again the particular case ${\bf \hat k} \parallel {\bf \hat x}$,
${\bf A} \parallel {\bf \hat y}$,
${\bf B} \parallel {\bf \hat z}$.  Taking $p_z = 0$ confines the motion
to the $xy$ plane so that phase space is five-dimensional (position
$(x,y)$, momentum $(p_x,p_y)$, and time $t$).
The constants (\ref{relgyro}) reduce the effective dimension of phase
space by two.  Choosing particular values $K_x = K_y = 0$
of the constants corresponds to shifting the gyrocenters of all possible
trajectories to the origin, removing two translational
degrees of freedom.  After this shift, $p_x$ and $p_y$ are no longer
independent coordinates but rather functions of $x$ and $y$ determined
by (\ref{relgyro}).

The existence of two constants of motion reduces the effective phase space
in this particular case from five dimensions to three.  Hamiltonian motion
in a two-dimensional phase space is always integrable, so that
three-dimensional motions such as the driven pendulum
and the Chirikov-Taylor problem~\cite{chirikov} are the simplest
that can exhibit nonintegrable
behavior.  The equations of
motion after a change to the independent
variable $\eta = t - x/c$ and introducing dimensionless
parameters $a_b = \omega_c / \omega$,
$a_w = e E_0 / m c \omega$, and a monochromatic wave $A(\eta)$ are
($\omega = m = c = 1$):
\begin{eqnarray}
\gamma &=& 1 + {{a_b}^2 y^2 + (a_b x - a_w \sin \eta)^2 \over
2 (1 + a_b y)} \cr
{d x \over d\eta} &=& {\gamma \over 1 + a_b y} - 1,\quad
{d y \over d\eta} = {a_b x - a_w \sin \eta \over 1 + a_b y}.
\label{scaled}
\end{eqnarray}
These equations are integrated numerically for various values of
$a_b$, $a_w$, and the initial conditions $x(\eta_0), y(\eta_0)$.

\begin{figure}
\epsfxsize=3.25truein
\centerline{\epsffile{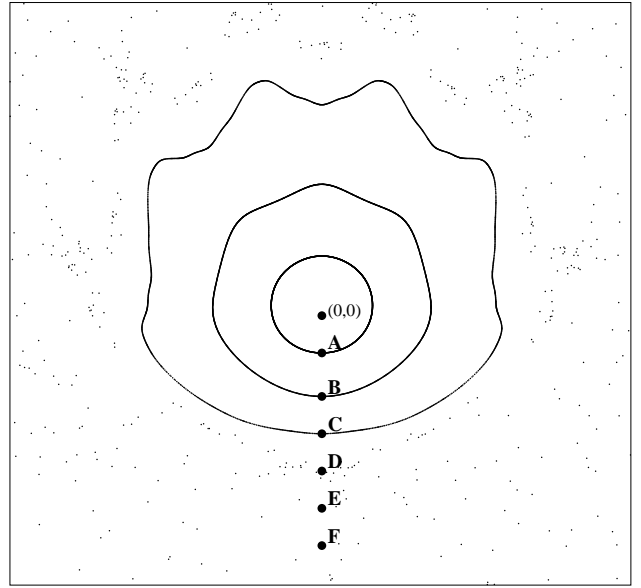}}
\caption{Surface of section showing trajectories from six different
initial conditions with $a_w = 0.1$, $a_b = 0.18.$}
\label{figtwo}
\end{figure}

\begin{figure}
\epsfxsize=3.25truein
\centerline{\epsffile{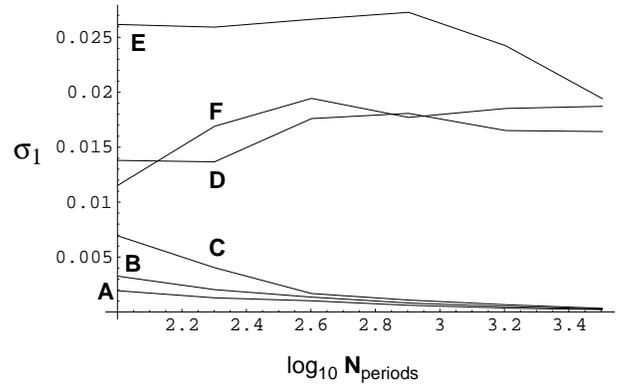}}
\caption{Numerical largest Liapunov exponent $\sigma_1$ calculated
at different times for trajectories starting on the six labeled
points in Fig. 2.}
\label{figthree}
\end{figure}

The equations of motion (\ref{scaled}) are periodic in
$\eta$ and plotting the electron's $(x,y)$ coordinates
after each period gives an area-preserving map of the plane to itself.
Fig.~\ref{figtwo} shows trajectories under this map.  For low
values of the electron initial energy, the motion is
quasiperiodic and nearly circular, while for large values of the initial
energy the motion is chaotic, as demonstrated by numerically calculated
Liapunov exponents (Fig.~\ref{figthree}).
As time increases the largest exponent for trajectories beginning
on points $D, E, F$ remains positive, indicating that neighboring
initial points separate exponentially rapidly~\cite{lichtenberg}.
The electron's initial energy affects the character of the motion because
$a_b$ can be much larger in the rest frame of an electron than in the lab frame
if the electron has a large initial velocity.  To eliminate this effect
the initial electron drift velocity is fixed at $0.5c$ henceforth.

The enhanced-mass description predicts that the guiding-center
of the electron moves in a circle: $(x_{\rm gc}(t),y_{\rm gc}(t)) =
(x_0 + (v_d / \Omega) \sin \Omega t, y_0 + (v_d / \Omega) \cos \Omega t),$
with $v_d = 0.5 c$ and $\Omega = {\omega_c / \gamma_0 \gamma_d} =
{a_b \omega \sqrt{1 - {v_d}^2} / \gamma_0}.$
At the end of each wave period, $x_{\rm gc}(t)$ and $y_{\rm gc}(t)$ are
compared with the actual location of the guiding-center calculated
from (\ref{scaled}).  As a dimensionless error measure
we use the normalized sum of squares error over a cyclotron orbit:
$E_{\rm gc} = ({\Omega / v_d})^2
% \sum_{\omega (t_n - x_n / c) = 2 n \pi}^{\Omega t_n < 2 \pi}
% {({\bf r}(t) - {\bf r}_{\rm gc}(t))^2}.
\sum {({\bf r}(t) - {\bf r}_{\rm gc}(t))^2}.$
The error is found to be quite small ($E_{\rm gc} < 0.01$) for
all values of $a_w$ studied as long as $a_b < 0.04.$
For each value of $a_w$, $E_{\rm gc}$ increases rapidly to order
unity once $a_b$ reaches a certain critical field: as a threshold
we define $a_b^{\rm crit}(a_w)$ as the value of $a_b$ where
$E_{\rm gc} = 0.01$.  Consistent with the predictions of the
enhanced-mass picture, $a_b^{\rm crit}$ remains nonzero for
large $a_w$ and in fact increases slightly with $a_w$.  Even
though high $a_w$ makes the equations (\ref{scaled}) quite nonlinear,
the motion remains quasiperiodic and nearly circular for
$a_b < a_b^{\rm crit}$.

\begin{figure}
\epsfxsize=3.25truein
\centerline{\epsffile{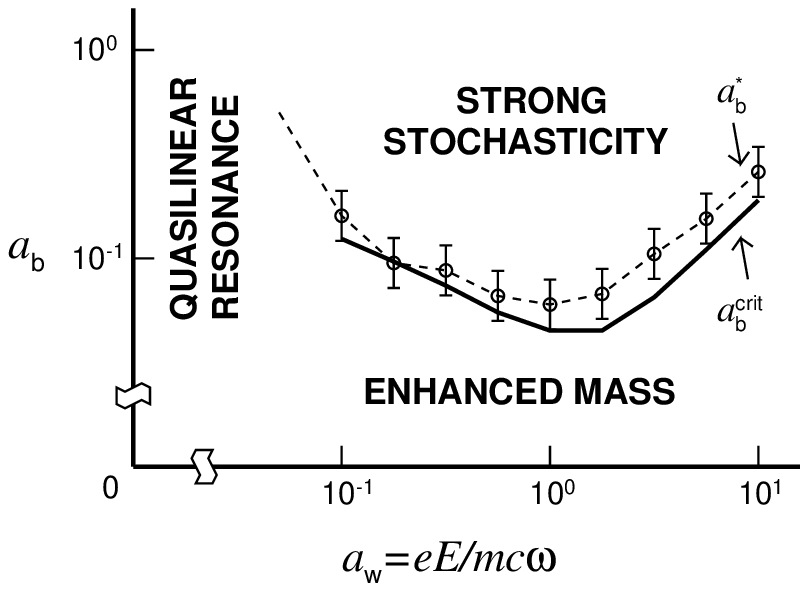}}
\caption{Numerical values $a_b^{\rm crit}$ and $a_b^*$ for
various $a_w$ as part of schematic phase diagram.  Error bars
are shown for $a_b^{*}$ because it is difficult to determine
precisely when the largest Liapunov exponent becomes positive.}
\label{figfour}
\end{figure}

As $a_b$ increases
above $a_b^{\rm crit}$,
the trajectory with initial velocity $0.5c$ becomes chaotic
(has a positive Liapunov exponent) at some value $a_b^*$.  For large
$a_w$, above $a_b^*$ the motion is strongly chaotic and the electron
energy fluctuates wildly.  This differs from nearly linear resonance
at small $a_w$ in that no tuning of frequencies is necessary for energy
gain and hence energy gain is not limited by relativistic detuning.
Fig.~\ref{figfour} shows numerical curves for
$a_b^{\rm crit}$ and $a_b^*$ as part of a schematic phase diagram.
The two curves are adjacent over four decades of beam intensity.
The dotted line (which was not calculated and is only schematic)
separating the quasilinear resonance phase from the strong stochasticity
phase can be defined by the destruction of the last invariant torus
at large energy, since in three dimensions such a torus bounds
the energy of trajectories contained within it.
Rax has previously proposed that
the stochastic motion of electrons in {\it multiple} plane waves
may give rise to high-energy cosmic rays~\cite{rax}.  The considerations
above indicate that a single plane wave, together with a sufficiently
strong magnetic field, is sufficient.

We verified that approximately the same boundary for the
guiding-center description applies when the wave contains two or three
frequencies.  It seems natural to conjecture that
the guiding-center region
in Fig.~\ref{figfour} also describes waves of finite bandwidth
and other orientations of the magnetic field (excluding the integrable
case ${\bf B} \parallel {\bf k}$).

The author wishes to thank Deepak Dhar for many helpful conversations.
This work was supported by a U.S. Fulbright grant and a fellowship from
the Hertz Foundation.

\end{document}